\documentclass[amsmath,amssymb,
aps,
nofootinbib,
prd,
superscriptaddress,
tightenlines,
notitlepage,
twocolumn,
showpacs,
floatfix]{revtex4-2}

\usepackage[utf8]{inputenc}
\usepackage{color}
\usepackage{xcolor}
\usepackage{hyperref}

\usepackage{graphicx}
\usepackage{dcolumn}
\usepackage{bm}

\newcommand{\DOA}{\affiliation{Department of Astronomy, School of Physics,
Peking University, Beijing 100871, China} }
\newcommand{\KIAA}{\affiliation{Kavli Institute for Astronomy and
Astrophysics, Peking University, Beijing 100871, China}}
\newcommand{\NAOC}{\affiliation{National Astronomical Observatories, Chinese Academy of Sciences, Beijing 100012, China}}

\begin{document}
\preprint{APS/123-QED}

\title{Axion induced spin effective couplings}
\author{Zihang Wang}\email{wzhax@pku.edu.cn}\DOA\KIAA

\author{Lijing Shao}\KIAA\NAOC

\date{\today}

\begin{abstract}
Detecting axionic dark matter induced electron or nucleon oscillating
electric dipole moment (OEDM) has become a new way for dark matter
searches. We re-examine such axion-spin couplings in external
electromagnetic fields. We point out that axion-photon interaction induces
an electron spin effective coupling, which is different from an OEDM. In
particular, the axion-spin effective coupling is directly related to
magnetic field rather than electric field. For axion-electron or
axion-nucleon couplings, an OEDM of fermion is introduced, whose effect in
ultralight axion cases depends on whether axion shift symmetry is manifest.
Specifically, ultralight axionic dark matter interactions that do not obey
the shift symmetry will be strongly constrained. We also extend the results
to the case where axion has a finite velocity.
\end{abstract}

\maketitle

\section{Introduction}

It has been known from various observations that most of the substance in
the Universe is invisible, in the form of dark matter and dark
energy~\cite{darkm,darke}. We observe dark matter only through
gravitational interaction, but we have never detected them directly yet.
Weakly interacting massive particles (WIMPs)~\cite{WIMP} are important
candidates for dark matter. Many proposed dark matter direct detection
experiments focus on detection of WIMPs, but after decades of efforts we
still have not found any convincing signal yet~\cite{WIMP2}. Thus other
kinds of dark matter candidates are getting more attention in recent years.

Another good dark matter candidate is the QCD axion, which was originally
proposed to solve the strong CP problem~\cite{ax1,ax2}. These theories
introduce a global U(1) symmetry, the so-called the Peccei-Quinn symmetry.
The symmetry is spontaneously broken below an energy scale $f_{a}$, and the
axion is the resulting pseudo-Goldstone boson. From the chiral perturbation
theory, the QCD axion parameters have a relation~\cite{precise},
\begin{equation}\label{eq:mass} m_{a}=5.70\, {\rm \mu
eV}\left(\frac{10^{12}\, {\rm GeV}}{f_{a}}\right)\, , \end{equation} where
$m_{a}$ is the axion mass. The relation holds for the QCD axion. In the
string theory~\cite{string} or other beyond standard model theories,
axion-like particles (ALPs) appear, whose properties are similar to the QCD
axion but do not necessarily satisfy Eq.~(\ref{eq:mass}). ALPs can have a
mass ranging from $\sim 10^{-33}\, {\rm eV}$ to $\rm keV$ scale or even
larger. Depending on the mass, these ALPs play very different roles in
astrophysics~\cite{marsh}. For example, an ALP with a mass $10^{-33}\, {\rm
eV}$ is a candidate of the dark energy~\cite{quintessence}. For an ALP with
a mass around $10^{-22}\, {\rm eV}$, it is a candidate of the fuzzy dark
matter~\cite{fuzzy}. For an ALP with a mass smaller than $10^{-10}\, {\rm
eV}$, it may help explain the possible TeV transparency of the
Universe~\cite{TeV}. The QCD axions with a mass around $10^{-5}\, {\rm eV}$
to $10^{-3}\, {\rm eV}$ are candidates of the cold dark matter~\cite{CDM}
and may contribute to stellar cooling~\cite{cooling}.

Ultralight axions refer to a type of ALPs with mass smaller than about
$10^{-18}\, \rm eV$~\cite{marsh}. In literature, ultralight axions have a
mass around $10^{-22}\, \rm eV$ are often discussed~\cite{fuzzy}, whose de
Broglie wavelength is about the size of a galaxy. The wave-like behavior of
ultralight axions may potentially solve the small scale crisis faced by
cold dark matter models (e.g., the core-cusp problem and missing-satellite
problem)~\cite{smallscale}.

Numerous methods have been proposed to probe axions or axionic dark matter,
including the microwave cavity experiments like ADMX~\cite{ADMX,ADMX2},
``light shining through wall'' experiments such as OSQAR~\cite{OSQAR}, and
Solar axion observations like CAST~\cite{CAST}. Although these experiments
have not found signal of axions, they have set constraints on axion
interactions. Recently XENON1T experiments~\cite{XENON1T} detected a
possible signal of Solar axions, but the results are still controversial.

Recently there are some new ideas of axionic dark matter detection. In
particular, searches for axion induced oscillating electric dipole moment
(OEDM) and axion induced spin precession become new ways for dark matter
searches~\cite{newobs}. CASPEr~\cite{CASPEr} is a proposed experiment to
detect nucleon OEDM using nuclear magnetic resonance. Detection of nucleon
OEDM using storage ring methods has also been proposed~\cite{storage}.
Hill~\cite{Hill2,Hill,Hill3,Hill4} proposed that axion-photon interaction
also introduces an OEDM of the electron, and the electric dipole radiation
induced by axionic dark matter is also discussed. Several experiments have
been proposed to detect the OEDM of the electron~\cite{eEDMde,eEDMde2}.
Similarly, axion-electron interaction may introduce an electron OEDM and
may be detected via spin precession~\cite{OEDM2}. Axion induced effects in
atoms and molecules are discussed in Refs.~\cite{molecular,molecular2}.
Besides, many techniques~\cite{electronEDM,electronEDM2} in constraining
electron static electric dipole moment (EDM) are helpful in axion detection
as well. Thus studying axion-spin interaction is crucial and timely.
Besides, there are extensions of these phenomena to macroscopic spin
interactions. For example, effects of spin-axion coupling in curved
spacetime that do not include electromagnetic fields are considered in
Ref.~\cite{spinaxion}.

In this paper, we will reconsider axion-spin couplings in applied electric
or magnetic field and find the effective interaction in non-relativistic
limit. We use the name ``axion'' but refer to general ALPs. We argue that
axion-photon interaction introduces an electron spin effective coupling,
which is not the same as OEDM. In particular, no physical effect occurs if
a constant electric field is applied, which agrees with the results in
Ref.~\cite{Hill}. We show that the spin effective interaction (for axions
with a zero velocity) is not directly related to the electric field but
related to the magnetic field. Because a time-varying electric field is
always accompanied with a magnetic field, a time-varying electric field
contributes to the spin effective interaction. A direct consequence of
these results is that experiments in Refs.~\cite{eEDMde,eEDMde2} can not be
used to detect axion-photon interaction in the zero axion velocity case.
Later in the paper we come to other axion interactions and include
corrections from a finite axion velocity. We will stress the importance of
the axion shift symmetry. For a type of ALP interaction that does not
satisfy the shift symmetry, strong constraints for ultralight axion
interactions are easily set by current experiments like static EDM
measurements. For those axion interactions that satisfy the axion shift
symmetry with only derivative axion couplings, the constraints from
experiments are much weaker in the ultralight axion case. The effective
spin interactions induced by such ultralight axion couplings are suppressed
by at least a factor of $m_{a}L\ll 1$. As long as the characteristic size
of experiments, $L$, is much smaller than the Compton wavelength of axions,
these effects will be suppressed.

The paper is arranged as follows. In Sec.~\ref{sec:photon}, we will discuss
the axion-photon interaction and its induced electron spin interactions. We
will stress its difference with an electron OEDM. In fact, the induced
electron spin interactions depend on magnetic field rather than electric
field. Hence experiments trying to detect such effects must be carefully
designed. Then we include axion velocity corrections. We will also extend
the result to the case with a time-varying external field. In
Sec.~\ref{sec:electron}, we will consider the axion-spin effective coupling
induced by the axion-electron interaction. We will consider two types of
effective axion-electron interactions found in literature, which turn out
to give different results. In Sec.~\ref{sec:neutron}, we consider effects
of axion-neutron interaction, which has been discussed in
Ref.~\cite{newobs}. We compare it with the axion-electron interaction case.
Finally, we come to an extended discussion in Sec.~\ref{sec:discussion}. In
this paper, Greek indices take values in $\{0, 1, 2, 3\}$ while Latin
indices take values in $\{1, 2, 3\}$.

\section{Effects of axion-photon interaction}
\label{sec:photon}

In the presence of axionic dark matter, the electron spin interacts
differently with the external electric or magnetic field, which may
manifest through OEDM or spin precession effects. We will first consider
the effective interaction induced by the axion-photon interaction. We will
neglect axion velocity and assume a static external electric and magnetic
field first in Sec.~\ref{subsec:zero:velocity}. Then in
Sec.~\ref{subsec:velocity} we extend the result to non-zero axion velocity
case. We will consider time-varying electromagnetic field in
Sec.~\ref{subsec:time:varying}.

Axion-photon interaction can be written as,
\begin{equation}
\mathcal{L}_{a\gamma\gamma}=\frac{1}{4}g_{a\gamma\gamma}aF_{\mu\nu}\tilde{F}^{\mu\nu}\, ,
\end{equation}
where $a$ is the axion field, $g_{a\gamma\gamma}$ is the axion-photon
coupling constant, $F_{\mu\nu} =
\partial_{\mu}A_{\nu}-\partial_{\nu}A_{\mu}$ is the electromagnetic field
tensor, and its dual $\tilde{F}_{\mu\nu}$ is defined as,
\begin{equation}
\tilde{F}_{\mu\nu}=\frac{1}{2}\epsilon_{\mu\nu\rho\sigma}F^{\rho\sigma}\, .
\end{equation}
Here we notice an important property,
\begin{equation}
F_{\mu\nu}\tilde{F}^{\mu\nu}=\partial_{\mu}\left(2\epsilon^{\mu\nu\rho\sigma}A_{\nu}\partial_{\rho}A_{\sigma}\right)\, .
\end{equation}
When we integrate by parts in action, we obtain,
\begin{equation}
S_{\rm int}=-\int d^{4}x \, \frac{1}{2}g_{a\gamma\gamma}(\partial_{\mu}a(x))\epsilon^{\mu\nu\rho\sigma}A_{\nu}(x)\partial_{\rho}A_{\sigma}(x)\, .
\end{equation}

The action above (except the axion mass term) satisfies a shift symmetry
$a\rightarrow a+a_0$ where $a_0$ is constant. As we will see, axion induced
spin couplings will be suppressed for small axion mass if axion
interactions obey the shift symmetry.

\subsection{Axions with zero velocity}
\label{subsec:zero:velocity}

It is well known that electron has a spin magnetic moment $\mu_{B}$, thus
electron spin interacts with external magnetic field. As we will see below,
due to the axion-photon interaction, the effective magnetic field seen from
the electron is slightly modified. Hence the presence of axionic dark
matter introduces an additional {\it effective} spin interactions.

\begin{figure}[htbp]
  \centering
  \includegraphics[width=8.6cm]{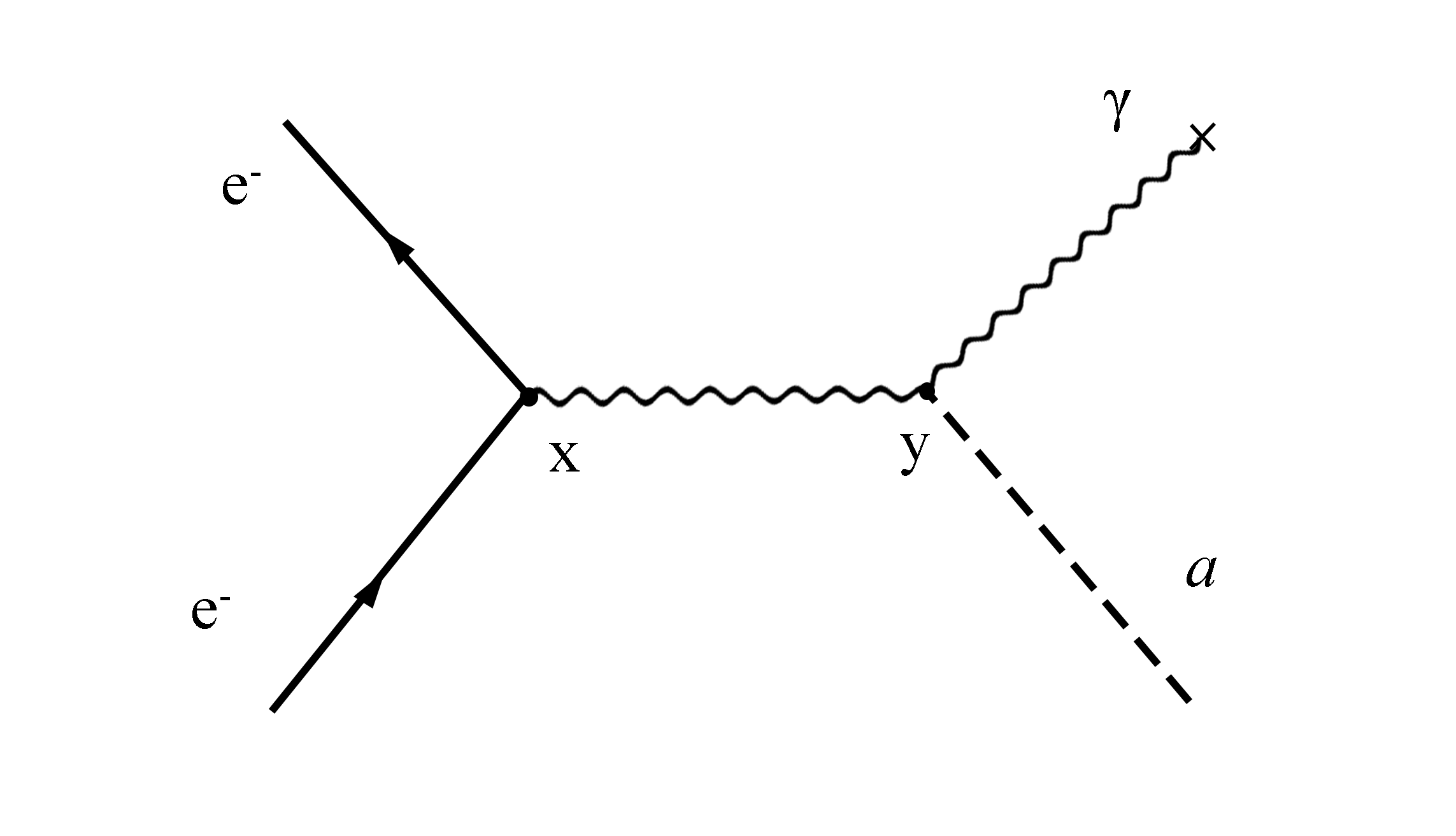}\\
  \caption{Feynman diagram that contributes to electron spin effective
  interactions, which is induced by axion-photon interaction. In the
  diagram, $a$ is the axion field, $x$ and $y$ are interaction points. The
  cross on the right represents interaction with external electromagnetic
  fields. } \label{img1}
\end{figure}

The main contribution comes from the Feynman diagram shown in
Fig.~\ref{img1}. We use the axion-photon interaction vertex
$(-g_{a\gamma\gamma}/2)\epsilon^{\mu\nu\rho\sigma}(\partial_{\mu}a)
A_{\nu}\partial_{\rho}A_{\sigma}$ and the magnetic moment interaction
$(-e/(2m_{e}))\bar{\psi}\sigma^{\mu\nu}\psi\partial_{\mu}A_{\nu}$. We use
the magnetic moment interaction because it is much easier for us to take
the non-relativistic limit. The result is the same as the usual
$-e\bar{\psi}\gamma^{\mu}\psi A_{\mu}$ interaction if we only consider spin
interactions of the electron. In fact, the axion-spin effective couplings
introduced by axion-photon interaction are only related to the magnetic
moment of the particle. We will calculate the amplitude in coordinate space
because it is easier to deal with the external field $A_{\mu}(x)$. The
amplitude is,
\begin{equation}
\begin{aligned}
\mathcal{M}&=-i\mu_{B}g_{a\gamma\gamma}\int d^{4}x d^{4}y \, \bar{\psi}(x)\sigma^{\mu\nu}\psi(x)\\& \times \left(\frac{\partial}{\partial x^{\mu}}G(x-y)\right)\epsilon_{\lambda\nu\rho\eta}\partial^{\lambda}a(y)\partial^{\rho}A^{\eta}(y)\, ,
\end{aligned}
\end{equation}
where $\mu_{B}=e/(2m_{e})$ is the Bohr magneton. We also define,
\begin{equation}\label{eq:G}
G(x-y)=-\int d^{4}q\, \frac{1}{q^{2}+i\epsilon}e^{iq\cdot(x-y)}\, .
\end{equation}
We then consider the non-relativistic limit of the electron,
\begin{equation}
\psi(x)=\left(
       \begin{array}{c}
         \chi(x) \\
         \varphi(x) \\
       \end{array}
     \right)e^{-im_{e}x^{0}}
\, .
\end{equation}
In the non-relativistic limit,
\begin{equation}
\bar{\psi}(x)\sigma^{ij}\psi(x)\rightarrow \chi^{\dag}(x)\epsilon^{ijk}\sigma^{k}\chi(x)\, ,
\end{equation}
\begin{equation}
\bar{\psi}(x)\sigma^{0i}\psi(x)\rightarrow 0\, .
\end{equation}
Hence the amplitude becomes,
\begin{equation}\label{eq:amp0}
\begin{aligned}
&\mathcal{M}=-i\mu_{B}g_{a\gamma\gamma}\int d^{4}x d^{4}y \, \psi^{\dag}(x)\sigma^{k}\psi(x) G(x-y) \times \\ &\left[\partial_{j}(\partial_{0}a(y)F^{kj}(y))  -\partial_{i}(\partial^{i}a(y)E^{k}(y)-\partial^{k}a(y)E^{i}(y))\right]\, ,
\end{aligned}
\end{equation}
where the electric field $E^{k}=F^{k0}$. We first consider the case that
axionic dark matter has velocity $v=0$, hence $\partial_{i}a=0$. In the
next subsection, we will come to velocity correction, which is suppressed
by a factor $v/c\sim 10^{-3}$. The only term that remains is,
\begin{equation}\label{eq:amp1}
\begin{aligned}
\mathcal{M}=i\mu_{B}g_{a\gamma\gamma}\int &d^{4}x d^{4}y \, \partial_{0}a(y^{0})\psi^{\dagger}(x)\sigma^{k}\psi(x)\\& \times G(x-y)(\nabla\times\vec{B})^{k}(y)\, ,
\end{aligned}
\end{equation}
In fact, $G(x-y)$ has a simple form in coordinate space. Eq.~(\ref{eq:G})
can be integrated directly. Define $t_{0}=x^{0}-y^{0}$,
$r=|\vec{x}-\vec{y}|$, and we obtain,
\begin{equation}
G(x-y)=-\frac{i}{4\pi^{2}}\frac{1}{t_{0}^{2}-r^{2}-i\epsilon}\, ,
\end{equation}
where $\epsilon$ is a small positive number. We use $\frac{1}{x-i\epsilon}={\cal P}\frac{1}{x}+i\pi\delta(x)$ to obtain,
\begin{equation}
G(x-y)=\frac{1}{4\pi}\delta(t_{0}^{2}-r^{2})-\frac{i}{4\pi^{2}}{\cal P}\frac{1}{t_{0}^{2}-r^{2}}\, ,
\end{equation}
where ${\cal P}\frac{1}{x}$ can be defined as
${\cal P}\frac{1}{x}=\frac{x}{x^{2}+\epsilon^{2}}$. We also
use,
\begin{equation}
\delta(t_{0}^{2}-r^{2})=\frac{1}{2r}\left[\delta(t_{0}-r)+\delta(t_{0}+r)\right]\, .
\end{equation}

Now we can integrate Eq.~(\ref{eq:amp1}) directly. We assume a coherent axion
background field $a(y^{0})=a_{0}e^{-im_{a}y^{0}}$.\footnote{Note that the
axion field is actually the real part of $a(y^0)$.} We also use the result,
\begin{equation}
\int dy^{0}{\cal P}\frac{e^{-im_{a}y^{0}}}{y^{0}}=-\pi i\, .
\end{equation}
For static fields, $\nabla\times \vec{B}=\vec{J}$, where $\vec{J}$ is the
current density. The total contribution is,
\begin{equation}\label{eq:axph0}
\begin{aligned}
\mathcal{M}=\frac{i\mu_{B}g_{a\gamma\gamma}}{4\pi}&\int  d^{4}x \, (\partial_{0}a(x^{0}))\\& \times \psi^{\dagger}(x)\sigma^{k}\psi(x)\int d^{3}y \, e^{im_{a}r}\frac{J^{k}(\vec{y})}{r}\, .
\end{aligned}
\end{equation}
This is equivalent to an amplitude under an effective interaction,
\begin{equation}\label{eq:axph1}
H_{\rm {int}}=-\mu_{B}g_{a\gamma\gamma}\frac{\partial a}{\partial t}\int d^{3}\vec{y}\, \frac{\vec{\sigma}\cdot\vec{J}(\vec{y})}{4\pi r}e^{im_{a}r} \, .
\end{equation}

We can understand the axion-spin coupling induced by axion-photon
interaction as follows. If axionic dark matter exists, in the presence
of an external static magnetic field $\vec{B}_{0}$, an effective
electromagnetic field will be induced. Assume that $\vec{B}_{0}$ is
produced by a constant electric current $\vec{J}(\vec{x})$ and the axion
has zero velocity. The induced electric field and magnetic field can be
calculated from a perturbative calculation of field
equations~\cite{inducedB,inducedB2},
\begin{equation}
\vec{E}_{\rm {eff}}(\vec{x},t)=-\frac{g_{a\gamma\gamma}a(t)}{4\pi}\int d^{3}\vec{y}\, \frac{e^{im_{a}r}-1}{r}\nabla\times \vec{J}(\vec{y})\, ,
\end{equation}
\begin{equation}\label{eq:beff}
\vec{B}_{\rm {eff}}(\vec{x},t)=\frac{ig_{a\gamma\gamma}m_{a}a(t)}{4\pi}\int d^{3}\vec{y}\,\frac{e^{im_{a}r}}{r}\vec{J}(\vec{y})\, ,
\end{equation}
where we have used $a(t)=a_{0}e^{-im_{a}t}$. We find that the effective
Hamiltonian~(\ref{eq:axph1}) is,
\begin{equation}
H_{\rm {int}}=-\mu_{B}\vec{\sigma}\cdot\vec{B}_{\rm {eff}} \, .
\end{equation}
This can be interpreted as electron magnetic moment interacting with the
effective magnetic field. In the limit $m_{a}r\ll 1$, $\vec{B}_{\rm {eff}}$
becomes,
\begin{equation}\label{eq:effba}
\begin{aligned}
\vec{B}_{\rm {eff}}(\vec{x},t)&=\frac{ig_{a\gamma\gamma}m_{a}a(t)}{4\pi}\int d^{3}\vec{y}\, \frac{\vec{J}(\vec{y})}{r}\\&=ig_{a\gamma\gamma}m_{a}a(t)\vec{A}(\vec{x}) \, .
\end{aligned}
\end{equation}

We note that in the limit $m_{a}r\ll 1$, which is true in experiments that
search for ultralight axions, the effective Hamiltonian is proportional
to the magnetic vector potential,
\begin{equation}
\vec{A}(\vec{x})=\int d^{3}\vec{y}\,\frac{\vec{J}(\vec{y})}{4\pi r}
\, .
\end{equation}

\begin{figure}[htbp]
  \centering
  \includegraphics[width=8.6cm]{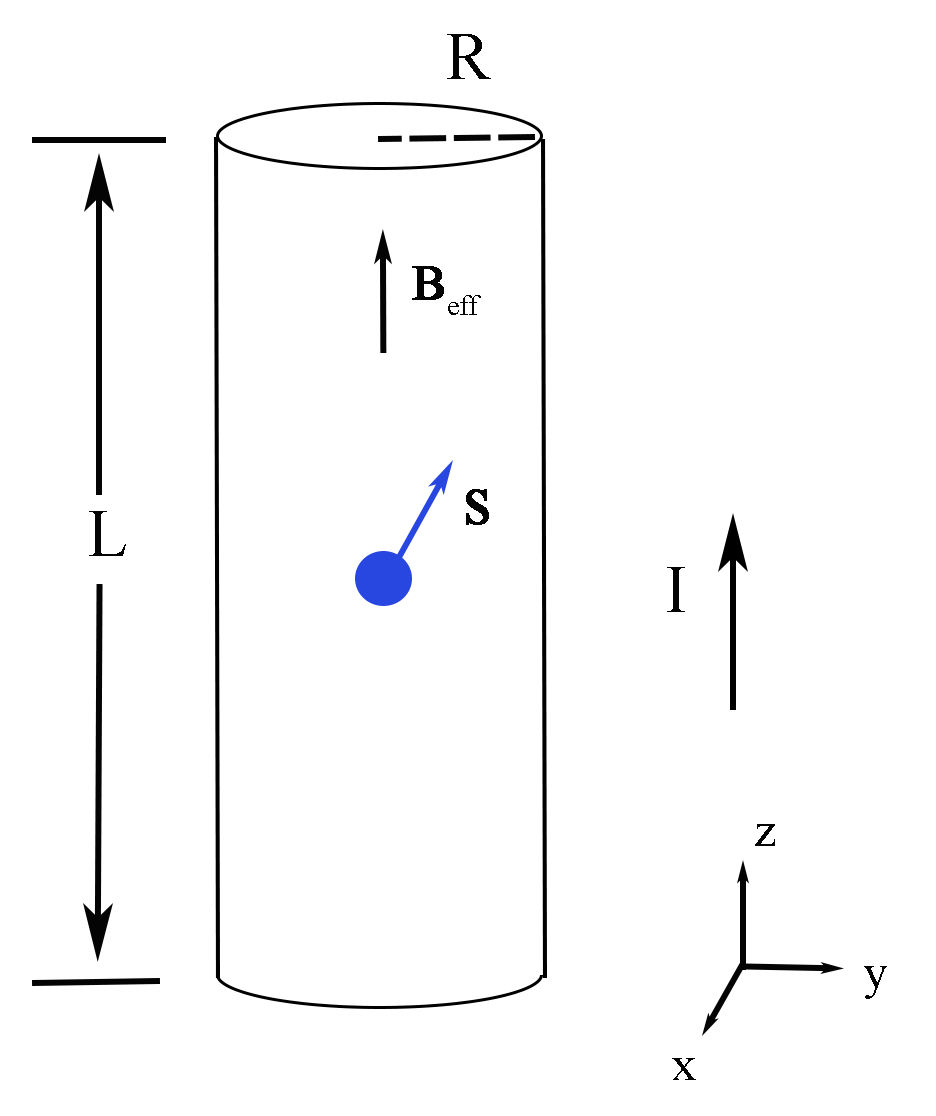}\\
  \caption{An example experimental setup to illustrate the effective
  interaction. A cylinder-shaped conductor with length $L$ and radius $R$
  is placed along $z$ axis. The conductor is hollow and an electric current
  flows in $+z$ direction in the surface of the conductor. Note that the
  magnetic field in the center of the cylinder is zero. But if axionic dark
  matter is present, an effective oscillating magnetic field is created in
  $z$ direction. Hence the spin of an electron (usually a molecular in
  experiments) in the center of the conductor will be affected and
  the spin precesses.}
  \label{img2}
\end{figure}

As an illustration of the effective interaction, consider an electron (or a
molecular in actual experiments) placed in the center of a cylinder shaped
conductor, as shown in Fig.~\ref{img2}. The conductor is hollow. An
electric current $I$ flows on the surface of the cylinder in $+z$
direction. The length of the conductor is $L$ and the radius of the
cylinder is $R$. If there is no axionic dark matter present, the magnetic
field in the center of the cylinder is zero. However, axionic dark matter
induces an effective magnetic field oscillating at a frequency
$\omega=m_{a}$. In the center of the cylinder the effective magnetic field
is,
\begin{equation}\label{eq:effb}
\vec{B}_{\rm {eff}}(t)=-\frac{g_{a\gamma\gamma}I\vec{e}_{z}}{4\pi}\frac{\partial a}{\partial t}\int_{-\frac{L}{2}}^{\frac{L}{2}} dz\, \frac{e^{im_{a}\sqrt{z^2+R^{2}}}}{\sqrt{z^2+R^{2}}}
\, .
\end{equation}
For ultralight axions with $m_{a}L\ll 1$ and $m_{a}R\ll 1$, the results are
easily obtained,
\begin{equation}
\vec{B}_{\rm {eff}}(t)=-\frac{g_{a\gamma\gamma}I\vec{e}_{z}}{4\pi}\frac{\partial a}{\partial t}\ln{\frac{L+\sqrt{L^2+4R^{2}}}{-L+\sqrt{L^2+4R^{2}}}}
\, .
\end{equation}
If we further assume that $R\ll L$, the results become,
\begin{equation}\label{eq:effmag}
\vec{B}_{\rm {eff}}(t)=-\frac{g_{a\gamma\gamma}I\vec{e}_{z}}{2\pi}\frac{\partial a}{\partial t}\ln{\frac{L}{R}}
\, .
\end{equation}

\begin{figure}[htbp]
  \centering
  \includegraphics[width=9cm]{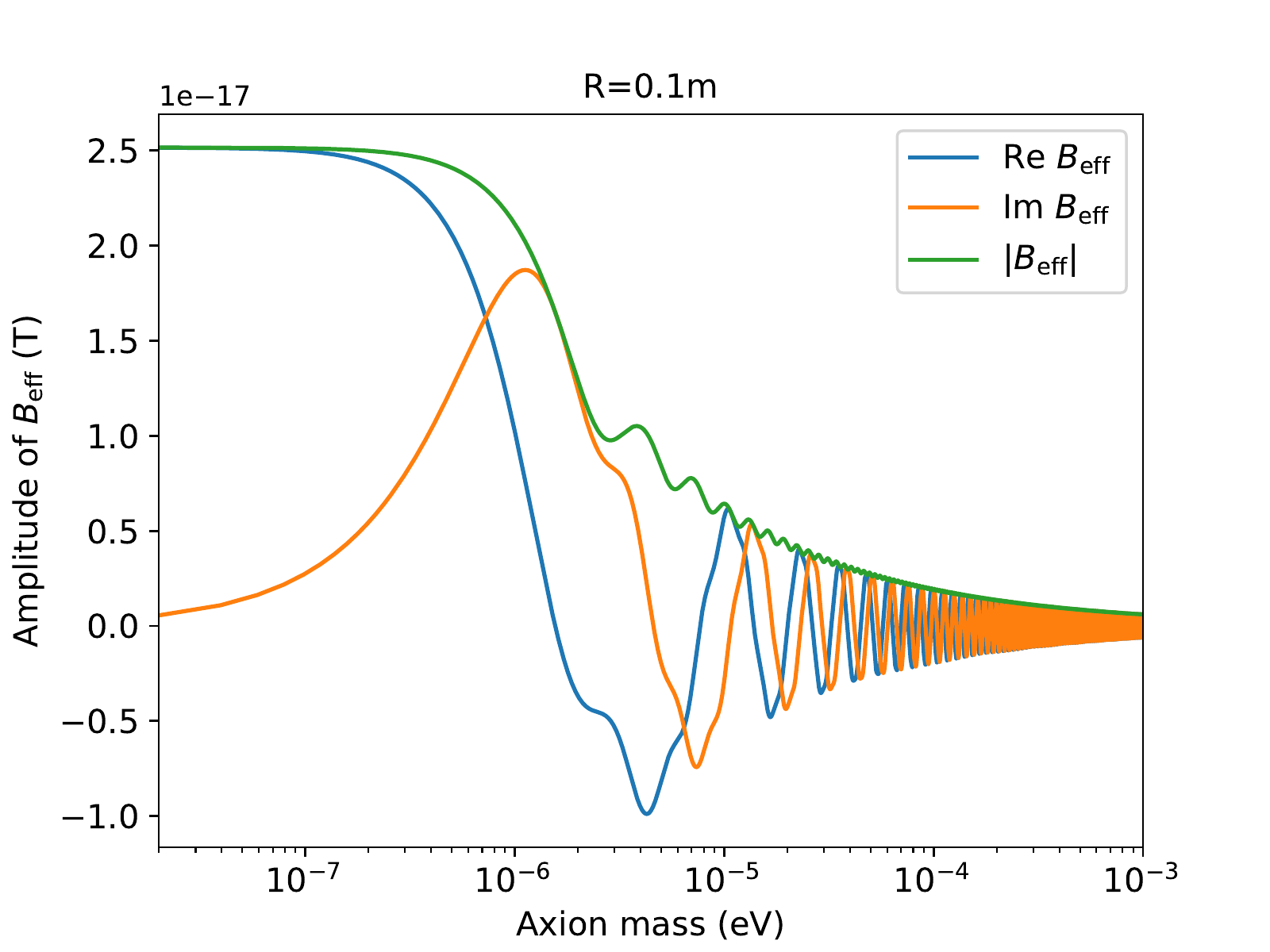}\\
  \caption{Effective magnetic field obtained from Eq.~(\ref{eq:effb}) in
  the center of the cylinder-shaped conductor in Fig.~\ref{img2} for
  different axion masses. We have fixed $L/R=10$ and $R=0.1\,{\rm m}$. We
  take $g_{a\gamma\gamma}=10^{-12}\,{\rm GeV^{-1}}$ and the dark matter
  density $\rho_{\rm DM}=0.3\,{\rm GeV/cm^{3}}$. The electric current is
  taken as $I=2\pi RB$ with $B=10\,{\rm T}$. Note that the effective
  magnetic field oscillates periodically with frequency $\omega \simeq
  m_{a}$, and the green line represents its amplitude $|\vec{B}_{\rm
  eff}|$. The blue and orange lines represent the real and the imaginary
  parts of $\vec{B}_{\rm eff}$. In the case that the size of the conductor
  is smaller than the Compton wavelength of axions, the amplitude of the
  effective magnetic field approaches a constant.}
  \label{img3}
\end{figure}

In Fig.~\ref{img3} we plot the amplitude of effective magnetic field in the
center of the conductor shown in Fig.~\ref{img2} for different axion masses.
We take $L/R=10$, $R=0.1\,{\rm m}$ and $g_{a\gamma\gamma}=10^{-12}\,{\rm
GeV^{-1}}$. The electric current is taken as $I=2\pi RB$ with $B=10\,{\rm
T}$. Axion mass $m_a = 10^{-6}\,{\rm eV}$ corresponds to a Compton
wavelength $\lambda_a = 1.2\, \rm m$. It is obvious that if the spatial
size of the conductor is much smaller than the axion Compton wavelength,
the amplitude of effective magnetic field approaches a constant value.
While for a larger axion mass, the amplitude of effective magnetic field
becomes smaller.

As an example, we will compare the effective interaction to the sensitivity
of electron anomalous magnetic moment measurements. We adopt a magnetic
field $B\sim10\,\rm T$, and use a characteristic magnetic field $B=I/(2\pi
R)$ to estimate the effective magnetic field in Eq.~(\ref{eq:effmag}). Note
that the characteristic magnetic field is not the magnetic field in the
center of the cylinder, which is zero. The effective interaction can be
regarded as a modification to the electron magnetic moment in such cases,
but we stress that it is intrinsically different from a magnetic moment.
For an electron whose spin is aligned in $z$ direction, the effective Hamiltonian is,
\begin{equation}
H_{\rm {eff}}(t)=\mu_{B}g_{a\gamma\gamma}BR\frac{\partial a}{\partial t}\ln{\frac{L}{R}}
\, .
\end{equation}
We thus find an electron effective magnetic moment,
\begin{equation}
|\mu_{\rm {eff}}(t)|=\mu_{B}g_{a\gamma\gamma}R\left|\frac{\partial a}{\partial t}\right|\ln{\frac{L}{R}}
\, .
\end{equation}
Take for example $L/R=10$, we obtain,
\begin{equation}\label{eq:amagmoment}
\begin{aligned}
\mu_{\rm {eff}}(t)&=2.5\times10^{-18}\mu_{B}\left(\frac{g_{a\gamma\gamma}}{10^{-12}\,{\rm GeV^{-1}}}\right)\left(\frac{R}{0.1\,{\rm m}}\right)\\& \times \left(\frac{\rho_{{\rm DM}}}{0.3\,{\rm GeV/cm^{3}}}\right)^{\frac{1}{2}}\sin{m_{a}t}
\, ,
\end{aligned}
\end{equation}
which is roughly five orders of magnitude below the experimental
sensitivity of anomalous magnetic moment measurements at present~\cite{partical}. The quantity in
Eq.~(\ref{eq:amagmoment}) is not an electron anomalous magnetic moment,
because $\mu_{\rm {eff}}$ does not couple to the actual magnetic field (but
couples to $B=I/(2\pi R)$) and it depends on the experimental
settings.

In many spin precession or dipole moment experiments, a uniform external
magnetic field is usually used. However, our calculation shows that the effects of axion-photon interaction on electron spin only appear for non-zero magnetic
vector potential in Eq.~(\ref{eq:effba}) when considering static external fields. This generally requires a
non-uniform magnetic field. For example, if the electric current is
arranged as a circle, and  electrons or molecules in experiment are
placed in the position which projects to the center of the circle, there
will be no detectable axion effects because the magnetic vector potential
is zero in that position.

We stress that the effective spin couplings induced by the axion-photon
interaction should not be regarded as an electron OEDM. For example, if
only a static electric field is applied, the magnetic field in
Eq.~(\ref{eq:amp1}) is zero and hence no axion-spin effective interaction
for zero-velocity axions. This behavior is in contrast with usual EDM or
OEDM. The result is consistent with the results in Ref.~\cite{Hill}, but
the author there incorrectly claimed that it is an electron OEDM. In fact, we
find the effective interaction is not related to the electric field but
related to $\nabla\times \vec{B}$. Hill~\cite{Hill} obtained a non-zero result
for a time-varying electric field because that the time-varying electric
fields are always accompanied with magnetic fields. Hill~\cite{Hill}
obtained an effective action,\footnote{Note that we have used a slightly
different definition of $g_{a\gamma\gamma}$ from Ref.~\cite{Hill}.}
\begin{equation}
\begin{aligned}
S&=\frac{1}{2}\mu_{B}g_{a\gamma\gamma}\int d^{4}x d^{4}y \, a(x)F^{\mu\nu}(x)\\&\times \left[S^{0}_{\mu\nu}(x)\delta^{4}(x-y)+\partial_{[\mu}G(x-y)\partial^{\lambda}S^{0}_{\nu]\lambda}(y)\right] \, ,
\end{aligned}
\end{equation}
where $S^{0}_{\mu\nu}=\bar{\psi}\sigma_{\mu\nu}\gamma^{5}\psi$. The first
term indeed has the form of an OEDM interaction in non-relativistic limit.
But the effective action also includes the second term, which is a
non-local interaction. When considering a non-relativistic electron interacting with a static electric field (no magnetic field), the second term exactly cancels the first term after integration. As we have shown,
the effective axion-spin coupling exists even if the electric field is
zero. Hence the point that axions induce an OEDM of the electron due to
axion-photon interaction is somehow misleading.

Chu et al.~\cite{eEDMde} proposed an experiment to detect axion-photon
interaction and use a constant electric field and a uniform magnetic field
to detect electron OEDM. However, our calculation shows that their
experiment can not detect axion-photon interaction if we neglect the axion
velocity.\footnote{Velocity effects are suppressed by
a factor of $v/c\sim 10^{-3}$.} As we will discuss in Sec.~\ref{sec:electron}, the experiment can detect axion-electron interaction in principle, but the effect may be much weaker than their expectations.

Another interesting fact is that the result is proportional to $m_{a}a(t)$
and hence the time derivative of axion field. As we have illustrated, the
axion-photon interaction satisfies axion shift symmetry, hence its physical
effect is proportional to the derivative of the axion field. Thus an extra
factor of $m_{a}$ appears, which suppresses the effective interaction for
ultralight axions. The density of axionic dark matter is given by,
\begin{equation}
\rho_{\rm DM}=\frac{1}{2}m_{a}^{2}a_{0}^{2}\, .
\end{equation}
Hence $a_{0}$ is proportional to $1/m_{a}$ for a fixed dark matter density.
The axion induced spin interaction Hamiltonian is proportional to $a_{0}$ and hence to $1/m_{a}$. If there were no axion-shift symmetry, we would have
concluded that $g_{a\gamma\gamma}$ can be strongly constrained in
experiments searching for ultralight axions. Due to the axion shift
symmetry, the effect of axion-interaction is proportional to time or
spatial derivatives of the axion field. Hence an extra $m_{a}$ appears,
cancelling the $1/m_{a}$ factor. Thus the effect on the spin precession
rate is not enhanced for a smaller axion mass. However, for those axion
interactions that do not obey the shift symmetry, for example axion-neutron
interaction which gives a neutron OEDM, the spin precession effects are
proportional to $1/m_{a}$. Such ALP interactions are strongly constrained
for ultralight axionic dark matter.

\subsection{Axion velocity effects}\label{subsec:velocity}

The motion of the Solar system in the Milky Way results in the relative
velocity of axionic dark matter as seen from the Earth. The axion velocity
is $v\sim10^{-3}c$ in terrestrial laboratories. If the axion has a non-zero velocity, we must keep the
spatial derivative of the axion field in Eq.~(\ref{eq:amp0}). The
contribution to the amplitude can be obtained using the same calculation
strategies,
\begin{equation}
\begin{aligned}
\mathcal{M}_{p}&=\frac{i\mu_{B}g_{a\gamma\gamma}}{4\pi}\int d^{4}x \, \psi^{\dag}(x)\sigma^{k}\psi(x)a(x)\int d^{3}\vec{y}\, \\& \times \frac{e^{-i\vec{p}\cdot\vec{r}}}{r}e^{im_{a}r} \left[\vec{p}^{2}E^{k}(\vec{y})-(\vec{p}\cdot\vec{E}(\vec{y}))p^{k}-\right.\\&\phantom{=\;\;}\left.ip^{i}\partial_{i}E^{k}(\vec{y})+ip^{k}\partial_{i}E^{i}(\vec{y})-m_{a}p^{i}F^{ki}(\vec{y})\right]\, ,
\end{aligned}
\end{equation}
where $\vec{p}$ is the axion momentum in the laboratory frame, and we have
defined $\vec{r}=\vec{x}-\vec{y}$ and $r=|\vec{x}-\vec{y}|$. We use $a(x)=a_{0}\exp(-ip\cdot x)$ to write derivatives of the axion field as $\vec{p}$ or $m_{a}$. Here we assume that the external electric and magnetic fields are static. We can write the effective Hamiltonian (including the zero-velocity part) as,
\begin{equation}
H_{\rm {int}}=-\mu_{B}\vec{\sigma}\cdot\left(\vec{B}_{\rm {eff}}+\vec{B}_{\rm {eff,p}}\right) \, ,
\end{equation}
where $\vec{B}_{\rm {eff}}$ is shown in Eq.~(\ref{eq:beff}) and
$\vec{B}_{\rm {eff,p}}$ is,
\begin{equation}\label{eq:velocityB}
\begin{aligned}
&\vec{B}_{\rm {eff,p}}(x)=\frac{g_{a\gamma\gamma}a(x)}{4\pi}\int d^{3}\vec{y}\, \frac{e^{-i\vec{p}\cdot\vec{r}}}{r}e^{im_{a}r}\\ &\times \left[\vec{p}^{2}\vec{E}(\vec{y})-(\vec{p}\cdot\vec{E}(\vec{y}))\vec{p}-i(\vec{p}\cdot\nabla)\vec{E}(\vec{y})\right.\\&\phantom{=\;\;}\left.+i\vec{p}(\nabla\cdot\vec{E}(\vec{y}))
+m_{a}\vec{p}\times\vec{B}(\vec{y})\right] \, .
\end{aligned}
\end{equation}
The results include three types of terms in the limit $m_{a}L\ll 1$, where
$L$ is the characteristic spatial size of the external electric or magnetic
field. The first two terms (after integrating over $\vec{y}$) are of order
$p^{2}L^{2}$. The third and the fourth terms are of order $pL$. The last
term is of order $m_{a}pL^{2}$. Here $m_{a}L$ is easily evaluated to be,
\begin{equation}
m_{a}L=5.07\left(\frac{m_{a}}{10^{-6}\, {\rm eV}}\right)\left(\frac{L}{\rm m}\right) \, .
\end{equation}
\begin{equation}
pL\approx10^{-3}m_{a}L \, .
\end{equation}
For ultralight axions whose mass is smaller than about $10^{-10}\, {\rm eV}$, we have $m_{a}L\ll 1$ and $pL\ll 1$ for $L\sim 1\, {\rm m}$. Hence we obtain
an approximate expression of the effective magnetic field for ultralight
axions,
\begin{equation}\label{eq:effbp}
\begin{aligned}
\vec{B}_{\rm {eff,p}}&=\frac{g_{a\gamma\gamma}a(x)}{4\pi}\int d^{3}\vec{y} \, \frac{1}{r} \\& \times \left[-i(\vec{p}\cdot\nabla)\vec{E}(\vec{y})+i\vec{p}(\nabla\cdot\vec{E}(\vec{y}))\right] \, .
\end{aligned}
\end{equation}
We note that the velocity effect appears only for a non-uniform external
electric field if $m_{a}L\ll 1$. For QCD axions with mass from
$10^{-5}\,{\rm eV}$ to $10^{-2}\,{\rm eV}$, however, all terms in
Eq.~(\ref{eq:velocityB}) are important. Thus for a uniform electric field,
the QCD axions induced spin interactions are non-zero.

We also note that the effective magnetic field depends on the direction of
axion velocity and external fields. Hence if such effects are detectable,
they will help us determine the motion of dark matter relative to the
Earth.

\subsection{Time-varying external fields}
\label{subsec:time:varying}

We can easily extend the result to the case where the external field is
time-varying. We will consider the case that the external electromagnetic
fields are changing periodically in time,
\begin{equation}
\vec{B}(\vec{x},t)=\vec{B}_{0}(\vec{x})e^{-i\omega t}\, ,
\end{equation}
where we have assumed that $\omega>0$. Here we will assume that axion
velocity is zero, and can be written as
$a(\vec{x},t)=a_{0}e^{-i(m_{a}t+\varphi)}/2+c.c.$. We use a slightly
different notation, which is necessary to obtain the correct result. As
shown in Sec.~\ref{subsec:velocity}, velocity effects are suppressed by at
least a factor of $10^{-3}$, and here we neglect them for simplicity.

The amplitude can be calculated as before, and the effective interaction
can be written as,
\begin{equation}
H_{\rm {int}}=-\mu_{B}\vec{\sigma}\cdot\vec{B}_{\rm {eff,t}} \, ,
\end{equation}
where
\begin{equation}\label{eq:effbt}
\begin{aligned}
&\vec{B}_{\rm {eff,t}}(\vec{x},t)=ig_{a\gamma\gamma}m_{a}a_{0}\int d^{3}y\, \frac{1}{8\pi r} \times \,\\&\left[e^{i(\omega+m_{a})(r-t)}e^{-i\varphi}-e^{i(\omega-m_{a})(r-t)}e^{i\varphi} \right](\nabla\times\vec{B}_{0}) \, .
\end{aligned}
\end{equation}
The effects are similar to the static external field case, but the effective magnetic field becomes a mixture of two frequencies $\omega+m_{a}$ and $|\omega-m_{a}|$. For ultralight axions with $m_{a}\ll \omega$, the two frequencies are almost equal. For the QCD axions and the external field frequency $\omega\sim m_{a}$, the first term in Eq.~(\ref{eq:effbt}) is a fast
oscillating term. The second term in Eq.~(\ref{eq:effbt}) is slowly
varying. Hence the first term can be neglected when considering processes
like electron spin precession. However, when considering axion induced
atomic transition processes, the two terms are both important.

\section{Effects of axion-electron interaction}
\label{sec:electron}

Axionic dark matter may also interact directly with electrons. In the
following we will find out the effective electron spin interactions induced
by axions.

Here we must distinguish two types of axion-electron effective interaction,
namely $\mathcal{L}_{1}=-i\lambda a\bar{\psi}\gamma_{5}\psi$ and
$\mathcal{L}_{2}=g_{aee}(\partial_{\mu}a)\bar{\psi}\gamma^{\mu}\gamma_{5}\psi$.
These two effective Lagrangian densities are frequently used in literature.
If we integrate by parts in the action and then use the Dirac equation, the
two interactions seem to be equivalent if $g_{aee}=\lambda/(2m_{e})$.
Indeed, in the Feynman diagram if the two electron lines connected to the
axion line are on shell, $\mathcal{L}_{1}$ and $\mathcal{L}_{2}$ are the
same. But more generally the two interactions are different and may lead to
different physical effects. Here we note that $\mathcal{L}_{1}$ does not
satisfy the axion shift symmetry while $\mathcal{L}_{2}$ does. For general
ALPs, we do not have reasons to impose axion shift symmetry. But for those
ALPs related to a spontaneously broken global U(1) symmetry like the QCD
axion, the shift symmetry should be satisfied. The most important result we
obtained in the following is that the parameter $\lambda$ can be strongly
constrained from spin-precession or OEDM experiments for ultralight axions
while the constraints for $g_{aee}$ are much weaker.\footnote{Here we
regard $\lambda$ and $g_{aee}$ independent to the axion mass.}

In Ref.~\cite{OEDM2}, the interaction Lagrangian $\mathcal{L}_{1}=-i\lambda
a\bar{\psi}\gamma_{5}\psi$ is used to derive the electron OEDM. Our result
for the electron OEDM differs from their result by a factor of $1/4$, but
the effective OEDM interaction Hamiltonian is the same. There could
be a typo in their final result. In the following, we will consider
$\mathcal{L}_{1}$ and $\mathcal{L}_{2}$ separately.

We will derive the results through field equations in the following, which
is more straightforward. The same results can also be carefully obtained
through Feynman diagrams. We take $\mathcal{L}_{1}=-i\lambda
a\bar{\psi}\gamma_{5}\psi$ first. Field equation can be derived from the
Lagrangian density,
\begin{equation}
\begin{aligned}
\mathcal{L}&=-\frac{1}{4}F_{\mu\nu}F^{\mu\nu}+\frac{1}{2}\partial_{\mu}a\partial^{\mu}a-\frac{1}{2}m_{a}^{2}a^{2}+\\&\bar{\psi}(i\gamma^{\mu}D_{\mu}-m_{e})\psi-i\lambda a\bar{\psi}\gamma_{5}\psi\, .
\end{aligned}
\end{equation}
We write the four-component spinor $\psi$ as a pair of two-component spinors $\chi$ and $\varphi$,
\begin{equation}\label{eq:wi}
\psi(x)=\left(
       \begin{array}{c}
         \chi(x) \\
         \varphi(x) \\
       \end{array}
     \right)e^{-im_{e}t}
\, ,
\end{equation}
and we obtain,
\begin{align}
(i\partial_{0}-eA^{0})\chi+(i\sigma^{i}\partial_{i}+e\sigma^{i}A^{i}-i\lambda a)\varphi&=0\, , \\
(-i\sigma^{i}\partial_{i}-e\sigma^{i}A^{i}-i\lambda a)\chi+(-i\partial_{0}-2m_{e}+eA^{0})\varphi&=0\, .
\end{align}
In the non-relativistic limit we have $i\partial_{0}\chi\ll2m_{e}\chi$,
$i\partial_{0}\varphi\ll2m_{e}\varphi$ and $eA^{0}\ll2m_{e}$. We obtain an
approximate equation in the non-relativistic limit,
\begin{equation}\label{eq:NR1}
\begin{aligned}
i\partial_{0}\chi&=eA^{0}\chi-(i\sigma^{i}\partial_{i}+e\sigma^{i}A^{i}-i\lambda a)\frac{1}{2m_{e}}\\& \times \left(1-\frac{i\partial_{0}}{2m_{e}}+\frac{eA^{0}}{2m_{e}}\right)
(-i\sigma^{j}\partial_{j}-e\sigma^{j}A^{j}-i\lambda a)\chi\, ,
\end{aligned}
\end{equation}
where the derivative in the bracket must act on all terms on the right. We
will only pick up those terms that involve the axion field, and neglect
higher order terms. The equation has the same form as the Schr\"{o}dinger
equation, and the term on the right hand side can be interpreted as an
effective Hamilton operator acting on $\chi$. There are terms proportional
to $\partial_{0}\chi$ on the right hand side in Eq.~(\ref{eq:NR1}). These
terms either cancel each other, or only give higher order contributions to
spin couplings. The electron is assumed to be static. If the axion has zero
velocity, the only term that remains is,
\begin{equation}
-\frac{e\lambda a}{4m_{e}^{2}}\sigma^{i}(\partial_{i}A^{0}+\partial_{0}A^{i})\chi=\frac{e\lambda a}{4m_{e}^{2}}\vec{\sigma}\cdot\vec{E}\chi\, ,
\end{equation}
where $\vec{E}$ is the electric field. This term contributes to an OEDM.
The magnitude of the electron OEDM is,
\begin{equation}
\begin{aligned}
d_{e,1}&=\frac{e\lambda a}{4m_{e}^{2}}=4.06\times10^{-31}e\cdot{\rm cm}\left(\frac{\rho_{{\rm DM}}}{0.3\,{\rm GeV/cm^{3}}}\right)^{\frac{1}{2}} \\
& \times \left(\frac{10^{-5}\,{\rm eV}}{m_{a}}\right)\left(\frac{\lambda}{10^{-16}}\right)
\cos(m_{a}t)\, ,
\end{aligned}
\end{equation}
Note that there is no axion shift symmetry here, so the OEDM is
proportional to the axion field rather than its derivatives. Thus the OEDM
obtained is inversely proportional to the axion mass. If ultralight axions
are dark matter, the OEDM can be very large in principle.

The oscillation period is,
\begin{equation}
\frac{2\pi}{m_{a}}=4.14\times10^{-10}\,{\rm s}\left(\frac{10^{-5}\,{\rm eV}}{m_{a}}\right) \, .
\end{equation}
For an axion lighter than about $10^{-17}\, {\rm eV}$, the period of OEDM
oscillation is large compared with typical experimental duration, and hence can be
regarded as a static EDM. For ultralight axions, $\lambda$ is strongly
constrained by electron EDM experiments.

If we allow a non-zero axion velocity, another term appears:
\begin{equation}\label{eq:v1}
-\frac{\lambda}{2m_{e}}\sigma^{i}(\partial_{i}a)\chi \, .
\end{equation}
This term is the same as the effective axion interaction term
$g_{aee}(\nabla a)\cdot\vec{\sigma}$, which has been studied in various
literature~\cite{newobs} and several experiments have been designed to
detect the interaction~\cite{molecular,precession}. Now we must know which
term dominates here. Take the axion velocity $v\sim10^{-3}c$,
we find that the OEDM term dominates if the electric field satisfies,
\begin{equation}
|\vec{E}|>5.2\times10^{9}\, {\rm V/m} \left(\frac{m_{a}}{\rm eV} \right) \, .
\end{equation}
This condition is easily satisfied in experiments for an axion mass smaller
than about $10^{-5}\, {\rm eV}$. The result above is consistent with
Ref.~\cite{OEDM2}, which has used $\mathcal{L}_{1}$ to obtain the results.

We next consider the effective interaction
$\mathcal{L}_{2}=g_{aee}(\partial_{\mu}a)\bar{\psi}\gamma^{\mu}\gamma_{5}\psi$.
The axion shift symmetry is satisfied here. Field equations can be
similarly derived,
\begin{equation}
\begin{aligned}
&\left(i\partial_{0}-eA^{0}+g_{aee}\sigma^{i}\partial_{i}a\right)\chi
\\ & +\left(i\sigma^{i}\partial_{i}+e\sigma^{i}A^{i}+g_{aee}\frac{\partial a}{\partial t}\right)\varphi=0\, ,
\end{aligned}
\end{equation}
\begin{equation}
\begin{aligned}
&\left(-i\sigma^{i}\partial_{i}-e\sigma^{i}A^{i}-g_{aee}\frac{\partial a}{\partial t}\right)\chi\\&+\left(-i\partial_{0}-2m_{e}+eA^{0}-g_{aee}\sigma^{i}\partial_{i}a\right)\varphi=0\, ,
\end{aligned}
\end{equation}
where we have used Eq.~(\ref{eq:wi}). In the non-relativistic limit, the
equation becomes,
\begin{equation}\label{eq:NR2}
\begin{aligned}
&i\partial_{0}\chi =\left(eA^{0}-g_{aee}\sigma^{m}\partial_{m}a\right)\chi-\\&
\left(i\sigma^{i}\partial_{i}+e\sigma^{i}A^{i}+g_{aee}\frac{\partial a}{\partial t}\right)\\ &\times\frac{1}{2m_{e}}\left(1-\frac{i\partial_{0}}{2m_{e}}+\frac{eA^{0}}{2m_{e}}-\frac{g_{aee}}{2m_{e}}\sigma^{j}\partial_{j}a\right)
\\&\times\left(-i\sigma^{k}\partial_{k}-e\sigma^{k}A^{k}-g_{aee}\frac{\partial a}{\partial t}\right)\chi\, .
\end{aligned}
\end{equation}
We neglect higher order terms and only keep those involving the axion
field. Here we will not consider the spatial motion of electrons, but this
can be important in experiments. We will take the external magnetic field $\vec{B}=0$ in
Eq.~(\ref{eq:NR2}) first. In Eq.~(\ref{eq:NR1}), terms proportional to $\partial_{0}\chi$ are cancelled to the lowest order. But in Eq.~(\ref{eq:NR2}), such terms do not cancel. To lowest order, there is an extra term on the right hand side:
\begin{equation}
-\frac{ig_{aee}}{2m_{e}^2}\frac{\partial a}{\partial t}\left(i\sigma^{i}\partial_{i}+e\sigma^{i}A^{i}\right) \partial_{0}\chi\, .
\end{equation}
Hence we must include corrections of this term by iterating $\partial_{0}\chi$ using Eq.~(\ref{eq:NR2}). This leads to many effective interaction terms. We only pick up two dominate terms here:
\begin{equation}\label{eq:v2}
-g_{aee}\sigma^{i}(\partial_{i}a) \chi\, ,
\end{equation}
\begin{equation}\label{eq:eOEDM}
\frac{ieg_{aee}}{4m_{e}^{2}}\frac{\partial a}{\partial t}\vec{\sigma}\cdot \vec{E} \chi\, .
\end{equation}
The second term contributes to an electron OEDM,
\begin{equation}
\begin{aligned}
d_{e,2}&=\frac{eg_{aee}m_{a}a}{4m_{e}^{2}}=4.06\times10^{-42}\, e\cdot{\rm cm}
\\ &\times \left(\frac{g_{aee}}{10^{-13}\,{\rm GeV^{-1}}}\right)\left(\frac{\rho_{{\rm DM}}}{0.3\,{\rm GeV/cm^{3}}}\right)^{\frac{1}{2}}\cos(m_{a}t)\, ,
\end{aligned}
\end{equation}
We note that $d_{e,2}\ll d_{e,1}$ for typical axion parameters with
$g_{aee}\sim \lambda/(2m_{e})$. Hence the axion shift symmetry suppresses
the induced electron OEDM. The second term is larger if the electric field
satisfies,
\begin{equation}\label{eq:E:value}
|\vec{E}|>5.3\times10^{15}\, {\rm V/m} \, .
\end{equation}
In electron EDM experiments, usually $\vec{E}$ is the effective electric
field in molecules. For fully polarized ThO molecule~\cite{electronEDM2},
the effective electric field is $78\, {\rm GV/cm}$, which is still much
smaller than $5.3\times10^{15}\, {\rm V/m}$ in Eq.~(\ref{eq:E:value}). Hence the first term dominates
in laboratory experiments.

In the following we will estimate the constraints on $\lambda$ and
$g_{aee}$ imposed by current electron EDM experiments. This should be
regarded as rough constraints and can be improved if we carefully adjust
the duration of ``time blocks"~\cite{electronEDM,electronEDM2}, within
which we measure the spin precession of molecules and average over these
measurements. In Ref.~\cite{electronEDM2} a block lasts for $T=60\, \rm s$,
which roughly means that we can only constrain ultralight axions with mass
$m_{a}\lesssim1/T$. For a higher mass the EDM effects may be averaged out. We again take $\rho_{\rm DM}=0.3\,{\rm GeV/cm^3}$ here.
Using the electron static EDM constraint~\cite{electronEDM2}
$d_{e}<1.1\times 10^{-29} \, e\cdot{\rm cm}$, the constraint for $\lambda$ is,
\begin{equation}
\lambda<2.71\times10^{-32}\left(\frac{m_{a}}{10^{-22}\,{\rm eV}}\right) \, , \quad {\rm for}~m_{a}\lesssim10^{-17}\,{\rm eV} \, .
\end{equation}
The constraint for $g_{aee}$ is,
\begin{equation}
g_{aee}<0.27\,{\rm GeV^{-1}} \, , \quad {\rm for}~m_{a}\lesssim10^{-17}\,{\rm eV} \, .
\end{equation}
It is obvious that the constraint for $g_{aee}$ is very weak compared with
stellar cooling constraint, $g_{aee}<2.8\times10^{-13} \, {\rm
GeV^{-1}}$~\cite{cooling2}. Hence for usual axions originated from a
spontaneously broken U(1) symmetry, electron static EDM does not give a
substantial constraint. The interaction $\mathcal{L}_{1}=-i\lambda
a\bar{\psi}\gamma_{5}\psi$ is not forbidden by symmetries for general ALPs,
in which case $\lambda$ is strongly constrained from static EDM experiments
for ultralight ALPs that compose the dark matter.

The experiments in Refs.~\cite{eEDMde,eEDMde2} were designed to detect
electron OEDM. As we have shown, axion-photon interaction does not
contribute to an OEDM but axion-electron interaction contributes. The
experiment can constrain $\lambda$, but it is difficult to give a
substantial constraint for $g_{aee}$ when considering OEDM interaction. This means that a large class of ALPs that satisfies the approximate shift symmetry is hard to detect in their experiment proposals. In contrast to the electron OEDM, the nucleon OEDM is much more likely to detect because the corresponding interaction does not satisfy the shift symmetry.

If we allow for a non-zero magnetic field, an additional effective interaction term appears:
\begin{equation}
-\frac{eg_{aee}}{2m_{e}^{2}}(\nabla a)\cdot \vec{B} \chi\, .
\end{equation}
This term is unrelated to the electron spin, and does not appear in the results of $\mathcal{L}_{1}$. It contributes an oscillating magnetic moment,
\begin{equation}
\vec{\mu}=\frac{eg_{aee}}{2m_{e}^{2}}(\nabla a)\, .
\end{equation}
This term is smaller than the OEDM interaction in Eq.~(\ref{eq:eOEDM}) because of the suppression of axion velocity $v\sim 10^{-3}$. The magnitude of the oscillating magnetic moment is,
\begin{equation}
\begin{aligned}
\mu=&4.2\times10^{-34}\mu_{B}\left(\frac{g_{aee}}{10^{-13}\,{\rm GeV^{-1}}}\right)\\ &\times\left(\frac{\rho_{\rm DM}}{0.3\,{\rm GeV/cm^{3}}}\right)^{\frac{1}{2}} \left(\frac{v}{10^{-3}}\right)\sin(m_{a}t)\, ,
\end{aligned}
\end{equation}
where the spatial dependence of the axion field contributes an unimportant phase for a static electron, which we neglect here. Such a small oscillating magnetic moment is not observable in experiments for near future.

\section{Effects of axion-neutron interaction}
\label{sec:neutron}

We finally briefly discuss axionic dark matter induced effective
interactions of the neutron spin, which is similar to the electron case.
The biggest difference is that there is an extra term that contributes to
the neutron OEDM~\cite{newobs} arising from the coupling
$(a/f_{a})G\tilde{G}$, where $G$ is the QCD field strength. Axion neutron
interactions include,
\begin{align}
\mathcal{L}_{n,1}&=g_{ann}(\partial_{\mu}a)\bar{\psi}\gamma^{\mu}\gamma_{5}\psi \, , \\
\mathcal{L}_{n,2}&=-\frac{i}{2}g_{d}a\bar{\psi}\sigma_{\mu\nu}\gamma_{5}\psi F^{\mu\nu} \, .
\end{align}
The first term has the same form as the axion-electron interaction, while
the second term contributes to an OEDM directly. In the non-relativistic limit,
the second term becomes,
\begin{equation}
g_{d}a\chi^{\dag}(x)\vec{\sigma}\chi(x)\cdot\vec{E}(x) \, ,
\end{equation}
where $\chi(x)$ is defined in Eq.~(\ref{eq:wi}). This term contributes to a
neutron OEDM,
\begin{equation}
\begin{aligned}
d_{n}&=g_{d}a=1.40\times10^{-30}\, e\cdot {\rm cm}\left(\frac{g_{d}}{10^{-10}\,{\rm GeV}^{-2}}\right)\\& \times \left(\frac{m_{a}}{10^{-5}\,{\rm eV}}\right)^{-1} \left(\frac{\rho_{\rm DM}}{0.3\,{\rm GeV/cm^{3}}}\right)^{\frac{1}{2}}\cos(m_{a}t)\, \, .
\end{aligned}
\end{equation}

The axion induced nucleon OEDM has been extensively studied before, and
many experiments are designed to detect it~\cite{CASPEr,storage}. We write
it here for completeness. We stress that the axion-nucleon
interaction $\mathcal{L}_{n,2}$ does not satisfy the axion shift symmetry due to instanton effects, and hence the neutron OEDM is proportional to $1/m_{a}$ for a fixed coupling constant $g_{d}$.\footnote{We regard $g_{d}$ as an
independent parameter here, not necessarily related to $m_{a}$.} The
constraint of $g_{d}$ from current static EDM experiments has been shown in
Ref.~\cite{nEDM}.

Besides, neutron has a magnetic moment
$\mu_{n}=-1.042\times10^{-3}\mu_{B}$~\cite{partical}. Hence axion-photon
interaction also introduces a coupling involving the neutron spin. But
because the magnetic moment of neutron is about $10^{-3}$ smaller than the
electron, the effect becomes difficult to detect. The reasoning is the
same as that in Sec.~\ref{sec:photon} with a replacement
$\mu_{B}\rightarrow\mu_{n}$.

\section{Discussion}
\label{sec:discussion}

If axions are the primary components of the dark matter, axion-spin
effective interactions will be introduced. These interactions may manifest
themselves in spin precession experiments or OEDM searches. Future
detectors with improved sensitivities may be able to detect axions via
these kinds of experiments.

We stress that axion-photon interaction effects on the electron spin is not
the same as an electron OEDM. The effective interaction is non-zero even if
the external electric field does not exist. The effect can be understood as
the electron spin interacting with an effective time-varying magnetic
field, whose frequency equals to the axion mass. The effective time-varying
magnetic field is independent of the electron and can be derived
through field equations. Hence such effects also appear for a macroscopic
magnetic moment, whose size is smaller than the Compton wavelength of the
axion. To detect such effects, we must use a spatially non-uniform magnetic
field, thus the effective magnetic field is non-zero.

When we discuss axion-photon interaction effects, we have assumed a
coherent axion background field, which can be written as
$a(x)=a_{0}e^{-ip\cdot x}$. However, if the scale of the external field is
larger than the coherent length $L_{0}\sim 10^{3}/m_{a}$, the results in
Sec.~\ref{sec:photon} do not apply because the axion field may have
different phase in different regions. In general, it is still possible that ALPs form a condensate and have a much larger coherent length than
$L_{0}$~\cite{condensate}.

For the axion-electron interaction, axion introduces an electron OEDM,
whose magnitude depends on the form of ALP-electron interaction.
Specifically, if the axion interaction obeys the shift symmetry as usual,
the electron OEDM is not enhanced for a very small axion mass. If the axion
interaction does not obey the shift symmetry, the electron OEDM is large
for ultralight axions. For an ultralight axion as dark matter with a mass
lighter than $10^{-17}\, {\rm eV}$, the oscillation period of OEDM is large
and it can be regarded as a static EDM in electron spin precession
experiments. Thus such interaction parameters including $\lambda$ and
$g_{d}$ are strongly constrained by current EDM experiments for ultralight
axionic dark matter.

It is possible that the dark matter density near the Earth or the Sun is
much larger than the local dark matter density $\rho_{\rm DM}=0.3\,{\rm
GeV/cm^{3}}$~\cite{axionhalo}. For example, ALPs may form miniclusters or
dilute axion stars~\cite{minicluster, Wang:2020zur}. If the Earth passes through such
high density region, the axion induced spin interactions will be easier to
detect. Another possibility is that some ALPs are trapped in the
gravitational potential of the Earth and the Sun, which may increase the
dark matter density by a factor of $10^{4}$~\cite{density}. This may
enhance the axion induced spin interaction by a factor $10^{2}$. But these
scenarios have large uncertainties and are still controversial.

\begin{acknowledgments}
We thank Li-Xin Li for helpful discussions.
Z. Wang was supported by the National Natural Science Foundation of China (Grant No.\ 11973014). L.~Shao was supported by the Young Elite Scientists Sponsorship Program by the China Association for Science and Technology (Grant No.\ 2018QNRC001), the National Natural Science Foundation of China (Grant Nos.\ 11975027 and 11991053), the National SKA Program of China (Grant No.\ 2020SKA0120300), and the Max Planck Partner Group Program funded by the Max Planck Society.
\end{acknowledgments}

\bibliography{refs}
\end{document}